\documentclass[11pt]{article}
\usepackage{amssymb,epsf,graphicx}
\usepackage{amsmath}
\allowdisplaybreaks[1]

\def\crbig{\\\noalign{\vspace {3mm}}}

\newcommand{\be}{\begin{equation}}
\newcommand{\ee}{\end{equation}}
\newcommand{\ben}{\begin{eqnarray}\displaystyle}
\newcommand{\een}{\end{eqnarray}}

\newcommand{\C}{{\cal C}}
\newcommand{\vac}{|0\rangle}

\newcommand{\sectiono}[1]{\section{#1}\setcounter{equation}{0}}

\begin{document}

{}~
\hfill\vbox{\hbox{hep-th/0609208}\hbox{SISSA 58/2006/EP}
\hbox{MIT-CTP-3773}}\break

\vskip 2.0cm

\centerline{\Large \bf The nonperturbative closed string tachyon vacuum to high level}

\vspace*{6.0ex}
\centerline{\large Nicolas Moeller
\footnote{E-mail: {\tt moeller@sissa.it}}}

\vspace*{1.5ex}

\centerline{\large \it International School for Advanced Studies (SISSA)}
\centerline{\large \it via Beirut 2-4, 34014 Trieste, Italy}

\vspace*{2.5ex}
\centerline{and}
\vspace{2.5ex}

\centerline{\large Haitang Yang
\footnote{E-mail: {\tt hyanga@mit.edu}}}

\vspace*{1.5ex}

\centerline{\large \it  Center for Theoretical Physics}
\centerline{\large \it Massachusetts Institute of Technology}
\centerline{\large \it Cambridge, MA 02139, USA}

\centerline{\&}

\centerline{\large \it Department of Applied Physics,}
\centerline{\large \it University of Electronic Science and Technology of China,}
\centerline{\large \it Chengdu, 610054, People's Republic of China}

\vspace*{10.0ex}

\centerline{\bf Abstract}
\bigskip

We compute the action of closed bosonic string field theory at quartic
order with fields up to level ten. After level four, the value of the
potential at the minimum starts oscillating around a nonzero negative
value, in contrast with the proposition made in \cite{vacuum}. We try
a different truncation scheme in which the value of the potential
converges faster with the level. By extrapolating these values, we are
able to give a rather precise value for the depth of the potential. 

\vfill \eject

\baselineskip=16pt

\vspace*{10.0ex}

\tableofcontents

\sectiono{Introduction and summary}

In this paper we are addressing the question whether closed bosonic
string theory has a stable vacuum. This is of course a
non-perturbative problem that needs to be approached in the context of
closed string field theory (CSFT) \cite{CSFT}. Its difficulty is
two-fold. Firstly, the action of CSFT is non-polynomial in the string
field. Secondly, the string field is composed of infinitely many
components. As an analytic solution of CSFT seems at present out of
reach (even in the light of the newly-discovered solution for the
vacuum of open string field theory \cite{Schnabl, Schnabl2}), we are
bound to numerical methods. The first difficulty is probably the most
serious but it is believed that truncating the action to a finite
power of the string field may furnish a good approximation. The second
difficulty is treated by level truncation, keeping in the string field
only component fields whose masses are not greater than a given level.

Until recently, only the quadratic and cubic terms of the CSFT action
could be computed. A level truncation calculation at this order was
done by Kosteleck\'y and Samuel in \cite{Kostelecky:1990mi}. They
truncated the string field to the massless level, keeping the tachyon,
graviton and auxiliary fields, and found a locally stable vacuum with
a positive tachyon expectation value. It is now understood
\cite{vacuum} that to cubic order we are missing some important
interactions, ones that can couple fields whose left-moving and
right-moving ghost numbers are not equal. The first scalar field
having this property is the ghost-dilaton which plays a central role.

In \cite{Belo}, Belopolsky endeavored the computation of the tachyon
effective potential up to quartic order. There were two terms to
calculate. Namely the contact term of four tachyons, and the Feynman
diagrams with two cubic vertices and four external tachyons. Those
terms were combined into one integral over the {\em whole} (i.e. not
reduced) moduli space of spheres with four punctures. Belopolsky then
found that this effective potential didn't have any local minimum,
the sign and magnitude of the quartic tachyon term were such as to
destroy the minimum existing at cubic order. There is however an
important flaw in the question of the tachyon effective potential
itself. As already mentioned, Yang and Zwiebach have shown in
\cite{vacuum}, that the zero-momentum ghost dilaton must be included
in the tachyon condensate as soon as we are considering quartic
terms. As this state is massless, it cannot be integrated out in
forming the tachyon effective potential. Instead one should consider
the effective potential of the tachyon {\em and} dilaton.

The computation of the quartic term in the CSFT action was made
possible in \cite{quartic}. This paper solves numerically the geometry
of the vertex and gives its solution in terms of fits which can be
used to calculate the coupling of any four states. The results of
\cite{quartic} were successfully checked in \cite{Yang:2005iu} by
verifying the cancellation of the effective coupling of marginal
fields to quartic order, and in \cite{dilaton} by checking the
cancellation of the effective term with four dilatons.

Yang and Zwiebach then proceeded in \cite{vacuum} to look for a
nonperturbative vacuum. This time the dilaton was taken proper care
of. They truncated the string field to level four, which included the
tachyon (level zero), the dilaton (level two) and four massive fields
at level four, and they found a stable vacuum with positive tachyon
and dilaton expectation values. The value of the potential at this
minimum is negative but seemed to approach zero as the level was
increased (and it is also shallower than the vacuum found with the
action truncated to cubic order). In the same paper, they studied the
low-energy effective action of the tachyon, dilaton and metric, and
found that a stable vacuum must have vanishing potential. They went on
to propose that this is valid for the full theory, and observed that
the numerical results seemed to confirm it. In such low-energy models,
a rolling tachyon solution is found. For a large class of potential,
the dilaton rolls to positive values corresponding to strong coupling
until the universe meets its fate in a big crunch \cite{Yang:2005rw}.
The natural interpretation of this vacuum would then be that all the
degrees of freedom of closed string theory have collapsed, in
particular the metric, and thus space-time, have disappeared. One
could then imagine that solitons of CSFT would correspond to
spacetimes of lower dimensionality. Some evidence that such solitons
exist in CSFT at quartic order was given in \cite{Bergman:2006pd}.
This interpretation is supported by open-closed $p$-adic string theory
\cite{Moeller:2003gg}.

In this paper, we continue the level truncation calculation of
\cite{vacuum} and push the computation to level ten. At this level,
the string field has a total of 158 fields and the computation of the
potential must be automatized. We use the symbolic calculator
Mathematica to perform antighost insertions, to calculate correlators
(and generate the conservation laws used to calculate them), and to
integrate the given results on the reduced moduli space using the
results of \cite{quartic}. The results for the nonperturbative vacuum
are not confirming the proposition \cite{vacuum} that its potential
should vanish. Instead we see that if we do level truncation in the
same way as in \cite{vacuum}, the depth of the potential oscillates
with the level, and the shallowness at level four is essentially an
illusion as the potential takes a dip at level six and then never
approaches zero as closely as it did at level four. We then use a
different truncation scheme, and find results that are consistent with
the former scheme but converge better. This leads us to conclude that
CSFT truncated to quartic order has a nonperturbative vacuum with a
nonzero potential, given by (\ref{limvacuum}).

We conclude this paper by asking how this result would change if we
include terms of higher order in the action. In \cite{quintic}, one of
us has solved numerically the geometry of the five-point vertex, and
checked the result with the dilaton theorem. At this time however only
the terms coupling five tachyons or five dilatons have been calculated
(other terms will be done in \cite{Moe-prog}). Although we should
really take terms at higher level as well, we are curious and look at
how our results change if we include the coupling of five tachyons. As
expected from the sign of this term, the potential at the vacuum is
pushed towards zero (but is still negative and nonzero). More
surprisingly, and perhaps hinting at something important, the
oscillations mentioned before are tamed.

\paragraph{}
The paper is structured as follows: In the rest of this section we
briefly summarize how to compute the quartic potential of CSFT. In
section \ref{s-conservation} we generalize the method of conservation
laws to compute correlators on the sphere with four punctures. We
describe our results of level truncation in Section \ref{s-results},
and finally we include the term with five tachyons and discuss our
results in Section \ref{s-conclusions}.

\paragraph{}
We shortly summarize how to calculate quartic multilinear functions,
more details can be found in \cite{dilaton, vacuum}. In our
conventions $\alpha'=2$, and the closed string field theory action is
\be S = -\frac{1}{\kappa^2} \left( \frac{1}{2} \langle\Psi| c_0^- Q_B
|\Psi\rangle + \frac{1}{3!} \left\{ \Psi,\Psi,\Psi \right\} +
\frac{1}{4!} \left\{ \Psi,\Psi,\Psi,\Psi \right\} + \ldots \right) \,,
\ee where $Q_B$ is the BRST operator, $c_0^\pm = \frac{1}{2} (c_0 \pm
\bar{c}_0)$, and $\left\{\ldots \right\}$ are the multilinear string
functions \cite{CSFT}. For the CSFT action to be consistent, the
string field $|\Psi\rangle$ must satisfy $(L_0 - \bar{L}_0)
|\Psi\rangle = 0$ and $(b_0 - \bar{b}_0) |\Psi\rangle = 0$.  We will
be working in the Siegel gauge $(b_0 + \bar{b}_0) |\Psi\rangle =
0$. As was shown in \cite{vacuum}, the minimal subspace of the Hilbert
space for the string field to live in when we are considering tachyon
condensation, is the one generated by the scalars obtained by
application on the vacuum of Virasoro, ghost and antighost
oscillators, and with the additional constraint $\Psi = -\Psi^\star$.
The action of $\star$ on a given state changes all left-moving
oscillators (Virasoro, ghost and antighost) into right-movers and
vice-versa, without changing their orders, and changes the factor in
front of the state by its complex conjugate.

To calculate the multilinear function of four states
$|\Psi_1\rangle, \ldots ,|\Psi_4\rangle$, one inserts them on the
sphere at the points $z=0$, $z=1$, $z=\infty$ and $z = \xi = x + y \,
i$, with an antighost insertion ${\cal B} {\cal B}^\star$, and one
then integrates the corresponding correlator over the reduced moduli
space of four-punctured spheres ${\cal V}_{0,4}$. It is reduced in the
sense that one excludes the spheres that can be obtained as Feynman
diagrams built with three-vertices. More explicitly
\be
\left\{ \Psi_1,\Psi_2,\Psi_3,\Psi_4\right\} = \frac{1}{\pi} 
\int_{{\cal V}_{0,4}} dx \wedge dy \langle \Sigma| 
{\cal B} {\cal B}^\star |\Psi_1\rangle |\Psi_2\rangle |\Psi_3\rangle |\Psi_4\rangle \,,
\label{multilinear}
\ee
where the antighost insertions are given by (\cite{dilaton})
\be
{\cal B} = \sum_{I=1}^4 \sum_{m=-1}^\infty \left(B_{m}^I b_m^{(I)} + 
\overline{C_{m}^I} \bar{b}_m^{(I)} \right) \quad , \quad 
{\cal B}^\star = \sum_{I=1}^4 \sum_{m=-1}^\infty \left(C_{m}^I b_m^{(I)} + 
\overline{B_{m}^I} \bar{b}_m^{(I)} \right) \,,
\label{BB}
\ee
whose coefficients $B_m^I$ and $C_m^I$ are determined by the four maps from the 
local coordinates $w_I$ to the uniformizer $z$
\ben
&& B_m^I = \oint \frac{dw}{2\pi i} \frac{1}{w^{m+2}} \frac{1}{h_I'} 
\frac{\partial h_I}{\partial \xi} \quad , \quad 
C_m^I = \oint \frac{dw}{2\pi i} \frac{1}{w^{m+2}} \frac{1}{h_I'} 
\frac{\partial h_I}{\partial \bar{\xi}} \label{Bim} \\ 
&& z = h_I(w_I; \xi, \bar{\xi}) = z_I(\xi, \bar{\xi}) + \rho_I(\xi, \bar{\xi}) \, w_I 
+ \sum_{n=2}^\infty{\alpha_{n,I}(\xi, \bar{\xi}) \, (\rho_I w_I)^n} \,.
\label{OmegaI}
\een
Note that for the puncture at infinity, we should use the coordinate
$t=1/z$ instead of $z$. Our notation here is a bit different from the
notation of \cite{dilaton, vacuum}. The $\beta_I$, $\gamma_I$ and
$\delta_I$ used there are related to $\alpha_{m,I}$ by 
\be
\beta_I \equiv \alpha_{2,I} \quad , \quad \gamma_I \equiv \alpha_{3,I} 
\quad , \quad \delta_I \equiv \alpha_{4,I} \,.
\label{bcd} \ee 
The $\alpha_{m,I}$ notation is more convenient at high level because
the computation of multilinear functions of fields of level $L$
requires $\alpha_{m,I}$ with $m=2,\ldots, L/2 +2$, in our case
$m=2,\ldots, 7$. These coefficients can be deduced from the quadratic
differential $\varphi = \phi(z) (dz)^2$ that gives the metric of the
interaction worldsheet. Namely it must have poles of second order with
residue minus one at the punctures, and its critical graph must be compact
(for more details see \cite{Strebel, Belo-Zwie, Belo, quartic}).  For
the four-vertex, it is given by
\be
\phi(z) = -\frac{(z^2-\xi)^2}{z^2(z-1)^2(z-\xi)^2} + 
\frac{a(\xi, \bar{\xi})}{z(z-1)(z-\xi)} \,.
\ee
The quadratic differential is thus determined by $a(\xi, \bar{\xi})$,
whose solution was constructed numerically in \cite{quartic}.  The
expressions of $\alpha_{m,I}$ follow by requiring that in the local
coordinates $w_I$, the quadratic differential takes the form
$\phi(w_I) = -1 / w_I^2$. All in all the integrand of
(\ref{multilinear}) can be expressed as an expression involving $\xi$,
$a$, $\partial a / \partial \xi$, $\partial a / \partial \bar{\xi}$,
and $\rho_I$, all of which can be directly estimated from the fits
given in \cite{quartic}, and correlators on the sphere. Our conventions for 
these correlators are the same as in \cite{dilaton, vacuum}, namely
\be
\langle c(z_1) c(z_2) c(z_3) \bar{c}(\bar{w_1}) \bar{c}(\bar{w_2}) 
\bar{c}(\bar{w_3}) \rangle = -2 \langle c(z_1) c(z_2) c(z_3) \rangle_o 
\cdot \langle \bar{c}(\bar{w_1}) \bar{c}(\bar{w_2}) 
\bar{c}(\bar{w_3}) \rangle_o \,,
\ee
and $\langle c(z_1) c(z_2) c(z_3) \rangle_o = (z_1-z_2) (z_1-z_3) (z_2-z_3)$ is 
the open string field theory correlator. These will be calculated with the help 
of the conservation laws described in Section \ref{s-conservation}

The way to do the integration in (\ref{multilinear}) was described in
\cite{dilaton}. The whole domain ${\cal V}_{0,4}$ can be decomposed
into six regions and their complex conjugates, such that
\be
\int_{{\cal V}_{0,4}} = \int_{{\cal A}} + \int_{\frac{1}{\cal A}} + 
\int_{1-{\cal A}} + \int_{\frac{1}{1-{\cal A}}} + 
\int_{1-\frac{1}{\cal A}} + \int_{\frac{{\cal A}}{{\cal A}-1}}
\ + \ \ {\rm complex \ conjugate} \,.
\label{integration}
\ee
All of these integrals can be expressed as integrals over ${\cal A}$
after pulling back their integrand (see \cite{dilaton} for more
details). And at last the two-dimensional region ${\cal A}$ was
described in \cite{quartic}, so we can do these integrals
numerically.

\sectiono{The conservation laws on the spheres with four punctures}
\label{s-conservation}

As outlined in the previous section, after we let the antighost
insertion ${\cal B} {\cal B}^\star$ act on the states, we must compute
correlators of the modified states (by which we mean the external
states modified by the antighost insertions). We could do that by
performing their conformal transformations from the local coordinates
to the sphere. But when the level increases it quickly becomes very
tedious to calculate the conformal transformations of the fields given
in terms of oscillators acting on the vacuum. We thus need an
alternative method for computing correlators; a very convenient one is
the method of conservation laws \cite{Ras-Zwi}. It was originally
constructed to calculate cubic interactions in Witten's cubic string
field theory, but it can be generalized to quartic interactions with
only notational complications. We review the main idea of this method
by considering, as an example, the conservation laws for the ghost
$c(z)$. We take a quadratic differential $\phi(z)$, so that the
product $\phi(z) c(z) dz$ transforms as a 1-form. And we consider a
small contour $\C$ on the sphere, which doesn't encircle any of the
punctures $0$, $1$, $\xi$ and $\infty$. If $\phi(z)$ is regular
everywhere, except possibly at the punctures, the contour can be
continuously deformed into the sum of four contours $\C_I$ around each
punctures. Expressing each integral in the local coordinates $w_I$, we
thus have
\be 
0 = \langle \Sigma | \sum_{I=1}^4\oint_{\C_I}\phi^{(I)}(w_I) c^{(I)}(w_I)
dw_I \,.
\ee
We have $c^{(I)}(w_I) = \sum_n{\frac{c^{(I)}_n}{w_I^{n-1}}}$, therefore
if $\phi^{(I)}(w_I)$ has a pole of order $n$, the $\C_I$ contour
integral will pick up an oscillator $c_{2-n}$ and oscillators with
higher indices. We can now explain the method: if we want to get rid
of an oscillator $c^{(I)}_{-n}$ at the puncture $I$, we choose a
$\phi(z)$ with a pole of order $2+n$ at the puncture $I$ and poles of
lesser order at the other punctures. We can then trade $c^{(I)}_{-n}$
for oscillators $c^{(J)}_{m}$ with $J=1,\ldots,4$ and
$m>-n$. Repeating this process, we will eventually be left with only
$c_{1}$'s.

\par
The conservation laws for the Virasoro oscillators are done much in the 
same way, except for the fact that $T(z)$ is not a tensor if the central 
charge is not zero. Under a conformal 
change of variable, it transforms as
\be
\tilde{T}(w) = \left(\frac{dz}{dw}\right)^2 T(z) + \frac{c}{12} S(z,w)\,,
\ee
where
\be
S(z,w) = \frac{z'''}{z'} - \frac{3}{2} \left( \frac{z''}{z'} \right)^2
\label{Schwartzian} \ee 
is the Schwartzian derivative (derivatives are with respect to $w$), 
and $c$ is the central charge. Now if $v(z)$ transforms like a vector field,
we see that the product $v(z)T(z)dz$ transforms as
\be
v(z) T(z) dz = \tilde{v}(w) \left(\tilde{T}(w) - \frac{c}{12} S(z,w) \right) dw \,.
\ee
Repeating the above idea of deforming a small contour, we find the Virasoro 
conservation laws
\be
\langle\Sigma| \sum_{I=1}^{4} \oint_{\C_I} v^{(I)}(w_I) \left(T^{(I)}(w_I) - 
\frac{c}{12} S(z, w_I)\right) dw_I = 0 \,.
\label{Lcons} \ee

\par
Since $b(z)$ has conformal weight two, it transforms as a stress-tensor with 
zero central charge, we can thus immediately deduce its conservation laws from 
(\ref{Lcons}).
\be
\langle\Sigma| \sum_{I=1}^{4} \oint_{\C_I} v^{(I)}(w_I) b^{(I)}(w_I) dw_I = 0 \,.
\label{bcons} \ee

\subsection{The first conservation laws for $T(z)$}

We compute here the first few conservation laws. The higher ones would
be too cumbersome to write down, but it will become clear that, like
the cubic ones, they can be easily generated on a computer. We start
by the conservation laws for $T(z)$, which are slightly easier than
$c(z)$ despite the presence of the central charge.  Before we begin we
must remark that in the case of the cubic vertex, due to its
cyclicity, one need only write the conservation laws for one puncture
(\cite{Ras-Zwi}). For example the conservation law to remove
$L_{-n}^{(1)}$ and the one to remove $L_{-n}^{(2)}$ are trivially
related by cycling the punctures $I \rightarrow I+1 \ ({\rm mod} \
3)$. For the quartic vertex there is no cyclic symmetry, and we have
to write the conservation laws for each of the four punctures.

\par 
Since we are considering only descendants of scalar fields with zero
momentum, which are annihilated by $L_{-1}$, we don't need the
conservation laws for $L_{-1}$. Should a $L_{-1}$ appear from another
conservation law, we can always commute it away. The first
conservation laws are thus the ones for $L_{-2}$, which we construct
now.

\par 
We start by expanding the Schwartzian derivative (\ref{Schwartzian})
in the local coordinates $w_I$ with the definitions (\ref{OmegaI}) and
(\ref{bcd}).
\be
S(z, w_I) = 6\rho_I^2\left(\gamma_I-\beta_I^2\right) + 
\rho_I^3 \left( 24 \delta_I-48\beta_I\gamma_I+24\beta_I^3 \right) \, w_I 
+ {\cal O}(w_I^2) \,.
\label{Schwarzian} \ee
The Schwartzian derivatives are regular, so they matter only where we 
have a pole. From the mode expansion
\be
T^{(I)}(w_I) = \sum{\frac{L_n^{(I)}}{z^{n+2}}}\,,
\ee
we see that we need a vector field $v(z)$ with a pole of order one at the 
puncture $I$, and regular everywhere else. In general we will denote  
$v_{n,I}(z)$ a vector field with expansion in the local 
coordinates $w_I$
\be 
v_{n,I}(w_I) = w_I^{-n+1} + {\cal O}(w_I^0) \,,
\ee 
and regular everywhere else. It can therefore be used to trade a $L^{(I)}_{-n}$ 
for oscillators $L^{(J)}_m$, $J=1,\ldots,4$, $m\geq -1$. It is easily seen 
recursively, that we can find such vectors for any $n \geq 2$. Indeed if we 
have $v_{m,I}(z)$ for $m < n$ and if we write the expansion of the vector field 
$u(z) = (z-z_I)^{-n+1}$ in the local coordinates $w_I$ as:
\be
u(w_I) = \sum_{m=-n+1}^{-1} a_m w_I^m + {\cal O}(w_I^0) \,,
\ee
we can take
\be
v_{n,I}(z) = \frac{1}{a_{-n+1}} \, \left( z^{-n+1} - \sum_{m=-n+2}^{-1} a_m 
v_{1-m,I}(z) \right) \,.
\ee
It will be useful to make the following definitions
\be
z_{IJ} \equiv z_I-z_J \,,
\qquad s_I \equiv \frac{1}{z_{IJ}} + \frac{1}{z_{IK}} \,,
\qquad q_I \equiv \frac{1}{z_{IJ} z_{IK}} \,,
\ee
where the set formed by $I$, $J$ and $K$ must be $\{1,2,3\}$ (regardless of order).
We are now ready to calculate the conservation laws. For $L_{-2}^{(I)}$ at the 
finite punctures $I=1,2,3$, we can take
\be
v_{2,I}(z) = \frac{\rho_I^2}{z_{IJ} z_{IK}} \frac{(z-z_J) (z-z_K)}{z-z_I} \,.
\ee 
Recalling that the local coordinates $w_I$ are related 
to the uniformizer $z$ (or $t=1/z$ for the puncture at infinity) 
through the conformal maps $h_I$, given by (\ref{OmegaI}) and (\ref{bcd}) 
and explicitly rewritten as
\ben
z &=& h_I(w_I) = z_I + \rho_I w_I + \rho_I^2 \beta_I w_I^2 + 
\rho_I^3 \gamma_I w_I^3 + \rho_I^4 \delta_I w_I^4 + \ldots \,, 
\quad I=1,2,3 \nonumber \\
t &=& h_4(w_4) = \rho_4 w_4 + \rho_4^2 \beta_4 w_4^2 + 
\rho_4^3 \gamma_4 w_4^3 + \rho_4^4 \delta_4 w_4^4 + \ldots \,,
\label{hI} \een 
and using the transformation law of a vector field
\be
\tilde{v}(w) = v(z) \frac{dw}{dz} \,,
\ee
we find the following expansions in the local coordinates $w_I$
\ben
v^{(I)}_{2,I}(w_I) &=& \frac{1}{w_I} + \rho_I \left(s_I-3\beta_I\right) + 
\rho_I^2 \left(-2 \beta_I s_I + q_I + 7 \beta_I^2 - 4 \gamma_I \right) \, w_I 
+ \ldots  
\nonumber \\
v^{(J)}_{2,I}(w_J) &=& \rho_I^2 \frac{z_{JK}}{z_{IJ}^2 z_{KI}} 
w_J + \ldots \,, \quad J\leq3 \,, \ J \neq I
\nonumber \\
v^{(4)}_{2,I}(w_4) &=& - \frac{\rho_I^2}{z_{IJ} z_{IK}} w_4 + \ldots \,.
\label{v2I} \een
For the puncture at infinity we take
\be
v_{2,4}(t) = \xi \frac{(t-1)\left(t-\frac{1}{\xi}\right)}{t} \rho_4^2 \,,
\ee
which has the expansions
\ben
v^{(4)}_{2,4}(w_4) &=& \frac{1}{w_4} - \rho_4 (1+\xi+3\beta_4) + 
\rho_4^2 \left(\xi+2\beta_4(1+\xi)+7\beta_4^2-4\gamma_4 \right) \, w_4 + 
\ldots
\nonumber \\
v^{(I)}_{2,4}(w_I) &=& -\rho_4^2 z_{IJ} z_{IK} \, w_I + 
\ldots \,, \quad I\leq3 \,.
\label{v24} \een
Now using (\ref{Schwarzian}), (\ref{v2I}) and (\ref{v24}) in (\ref{Lcons}) 
we find the conservation laws for $L_{-2}$
\ben
0 &=& \langle\Sigma| \left( L_{-2} + \frac{c}{2} \rho_I^2 (\beta_I^2-\gamma_I) 
+ \rho_I \left(s_I-3\beta_I\right)  \, L_{-1} + 
\rho_I^2 \left(-2 \beta_I s_I + q_I + 7 \beta_I^2 - 4 \gamma_I \right) \, L_0 + 
\ldots \right)^{(I)}
\nonumber \\
&+& \sum_{\genfrac{}{}{0pt}{}{J=1}{J\neq I}}^3 \langle\Sigma| 
\left( \rho_I^2 \frac{z_{JK}}{z_{IJ}^2 z_{KI}} 
L_0 + \ldots \right)^{(J)}
+ \langle\Sigma| \left( -\frac{\rho_I^2}{z_{IJ} z_{IK}} L_0 + 
\ldots \right)^{(4)} \,, \quad I=1,2,3
\nonumber \\
0 &=& \langle\Sigma| \left( L_{-2} + \frac{c}{2} \rho_4^2 (\beta_4^2-\gamma_4) 
- \rho_4 (1+\xi+3\beta_4) \, L_{-1} + 
\rho_4^2 \left(\xi+2\beta_4(1+\xi)+7\beta_4^2-4\gamma_4 \right) \, L_0 + 
\ldots\right)^{(4)}
\nonumber \\
&+& \sum_{I=1}^3 \langle\Sigma| \left( -\rho_4^2 z_{IJ} z_{IK} \, L_0 
+ \ldots \right)^{(I)} \,,
\label{CLL-2} \een
where the dots indicate oscillators with indices greater than zero.

\par
Now we go one step further and write the conservations laws for $L_{-3}$. 
We are again expanding them up to $L_0$, so, together with the laws for $L_{-2}$, 
they can be used to compute the matter part of all quartic correlators with 
one field of level six and three other fields of level up to four. We take
\ben
v_{3,I}(z) &=& \frac{\rho_I^3}{z_{IJ} z_{IK}} \frac{(z-z_J) (z-z_K)}{(z-z_I)^2}
- \rho_I (s_I-4\beta_I) \, v_{2,I}(z) \nonumber \\
v_{3,4}(t) &=& \rho_4^3 \xi \frac{(t-1)\left(t-\frac{1}{\xi}\right)}{t^2} + 
\rho_4 (1+\xi+4\beta_4) \, v_{2,4}(t) \,,
\een
from which we find the conservation laws
\ben
0 &=& \langle\Sigma| \left( L_{-3} - 
c \, \rho_I^3 \left( 2 \delta_I-4\beta_I\gamma_I+2\beta_I^3 \right)
+ \rho_I^2 \left(-\beta_I^2-5\gamma_I+4\beta_Is_I-s_I^2+q_I\right) \, L_{-1} 
\right.
\nonumber \\
&& \left. + \rho_I^3 \left(2\beta_I^3+12\beta_I\gamma_I-6\delta_I-8\beta_I^2s_I+2\beta_Iq_I 
+2\beta_Is_I^2-s_Iq_I \right) \, L_0 + \ldots \right)^{(I)}
\nonumber \\
&+& \sum_{\genfrac{}{}{0pt}{}{J=1}{J\neq I}}^3 \langle\Sigma| \left(
\frac{\rho_I^3z_{JK}}{Z_{IJ}^2z_{IK}} \left(\frac{1}{z_{IJ}} + s_I - 4\beta_I \right) \, 
L_0 + \ldots \right)^{(J)}
+ \langle\Sigma| \left( \frac{\rho_I^3}{z_{IJ} z_{IK}} (s_I-4\beta_I) \, L_0 
+ \ldots \right)^{(4)}
\nonumber \\
0 &=& \langle\Sigma| \left( L_{-3} - 
c \, \rho_4^3 \left( 2 \delta_4-4\beta_4\gamma_4+2\beta_4^3 \right)
- \rho_4^2 \left( \beta_4^2 + 4(1+\xi)\beta_4 + 5\gamma_4 +1+\xi+\xi^2 
\right) \, L_{-1} \right.
\nonumber \\
&& \left. + \rho_4^3 \, \left(2 \beta_4^3+8(1+\xi)\beta_4^2+2(1+3\xi+\xi^2)\beta_4 
+12\beta_4\gamma_4  - 6 \delta_4 + \xi+\xi^2 \right) \, 
L_0 + \ldots \right)^{(4)}
\nonumber \\
&+& \sum_{I=1}^3 \langle\Sigma| \left( 
-\rho_4^3 \left(z_I^2(1-\xi)(-1)^I + z_{IJ} z_{IK} (1+\xi+4\beta_4) \right) \,
L_0 + \ldots \right)^{(I)} \,.
\label{CLL-3} \een

We emphasize again that the conservation laws for $b_{-n}$ are the same 
as for $L_{-n}$ after setting the central charge $c$ to zero.

\subsection{The first conservation laws for $c(z)$}
If the string states are in the Siegel gauge, they will carry
no $c_0$ oscillators, so we don't need the conservation laws for
$c_0$. One may worry that a $c^{(I)}_0$ may arise from a term $w_I^{-2}$ in 
another conservation law, but we can avoid this because we
can always remove such a term by subtracting multiples of the 
quadratic differentials given by
\ben
\phi_{0,I}(z) &=& \frac{z_{IJ} z_{IK}}{(z-z_I)^2 (z-z_J) (z-z_K)} , 
\quad I = 1,2,3  \nonumber \\
\phi_{0,4}(t) &=& \frac{\xi^{-1}}{t^2 (t-1) (t-\xi^{-1})} \,.
\een
We see that $\phi_{0,I}(z)$ has a pole of order $2$ with unit coefficient 
at the puncture $z_I$, and poles of order one at two other punctures. For 
$I<4$, $\phi_{0,I}(z)$ is finite at infinity. We denote by 
$\phi_{n,I}(z)$ a quadratic differential with expansion in the local 
coordinates $w_I$
\be 
\phi_{n,I}(w_I) = w_I^{-n-2} + {\cal O}(w_I^{-1}) \,,
\ee 
and regular everywhere expect for possible poles of order one at 
other punctures. It can therefore be used to trade a $c^{(I)}_{-n}$ 
for oscillators $c^{(J)}_m$, $J=1\ldots,4$, $m\geq 1$. Again, 
it is easy to see that we can find such quadratic differentials 
for any $n \geq 1$.

\par
We can now write the conservation laws for $c_{-1}$. For the finite punctures 
$I=1,2,3$, we can take 
\be
\phi_{1,I}(z) = \left(\frac{\rho_I}{z-z_I} - \rho_I (\beta_I-s_I) \right)
\phi_{0,I}(z) \,.
\label{phi11} \ee
Using the transformation law of a quadratic differential $\phi(z)$
\be
\tilde{\phi}(w) dw^2 = \phi(z) dz^2 \,,
\ee
and the conformal maps (\ref{hI}), 
we can write the expansions of $\phi_{1,I}(z)$ in the local coordinates
\ben
\phi^{(I)}_{1,I}(w_I) &=& \frac{1}{w_I^3} + \rho_I^2 \left( 
-4 \beta_I^2 + \beta_I s_I + 3 \gamma_I - q_I \right) \frac{1}{w_I} + 
\ldots \nonumber \\
\phi^{(J)}_{1,I}(w_J) &=& - \frac{\rho_I \rho_J z_{IK}}{z_{IJ} z_{JK}} 
\left( \frac{1}{z_{IJ}} + \beta_I - s_I \right) \frac{1}{w_J} + \ldots 
\nonumber \\
\phi^{(4)}_{1,I}(w_4) &=& {\cal O}(w_4^0) \,.
\een
For the puncture at infinity we take
\be
\phi_{1,4}(t) = \frac{\rho_4}{t^3} - \beta_4 \rho_4 \phi_{0,4}(t) \,,
\ee
which has the expansions
\ben
\phi^{(4)}_{1,4}(w_4) &=& \frac{1}{w_4^3} + \rho_4^2 
\left(3\gamma_4-4\beta_4^2-(1+\xi)\beta_4 \right) \frac{1}{w_4}
+ \ldots \nonumber \\
\phi^{(I)}_{1,4}(w_I) &=&  \rho_4 \rho_I \left( \delta_{I1} - (-1)^I 
(1-\delta_{I1}) \frac{\beta_4}{1-\xi} \right) \frac{1}{w_I} + \ldots \,.
\een
From these expansions we deduce the conservation laws for $c_{-1}$
\ben
0 &=& \langle\Sigma| \left( c_{-1} + 
\rho_I^2 \left( -4\beta_I^2+\beta_I s_I + 3\gamma_I-q_I\right) \, c_1 + 
\ldots \right)^{(I)}
\nonumber \\
&+& \sum_{\genfrac{}{}{0pt}{}{J=1}{J \neq I}}^3 \langle\Sigma| \left( 
-\frac{\rho_I\rho_J z_{IK}}{z_{IJ} z_{JK}} \left( \frac{1}{z_{IJ}}+\beta_I-s_I \right) 
\, c_1 + \ldots \right)^{(J)} + 
\langle\Sigma| \left( \ldots \right)^{(4)}
\nonumber \\
0 &=& \langle\Sigma| \left( c_{-1} + 
\rho_4^2 \left(3\gamma_4-4\beta_4^2-(1+\xi)\beta_4\right) c_1 + 
\ldots \right)^{(4)}
\nonumber \\
&+& \sum_{I=1}^3 \langle\Sigma| \left( \rho_4 \rho_I \left(\delta_{I1} - 
(-1)^I (1-\delta_{I1}) \frac{\beta_4}{1-\xi} \right) \, c_1 + 
\ldots \right)^{(I)} \,.
\label{CLc-1} \een
We now write the next conservation laws, for $c_{-2}$. For the 
vector fields, we take
\ben
\phi_{2,I}(z) &=& \frac{\rho_I^2}{(z-z_I)^4} + 
2 \rho_I^2 (\beta_I^2-\gamma_I) \, \phi_{0,I}(z) 
\nonumber \\
\phi_{2,4}(t) &=& \frac{\rho_4^2}{t^4} + 
2 \rho_4^2 (\beta_4^2-\gamma_4) \, \phi_{0,4}(t) \,.
\een
And we find 
\ben
0 &=& \langle\Sigma| \left( c_{-2} + 
2 \rho_I^3 \left( 4\beta_I^3-6\beta_I\gamma_I+2\delta_I+
(\gamma_I-\beta_I^2)s_I \right) \, c_1 + 
\ldots \right)^{(I)}
\nonumber \\
&+& \sum_{\genfrac{}{}{0pt}{}{J=1}{J \neq I}}^3 \langle\Sigma| \left( 
2 \rho_I^2 \rho_J (\beta_I^2-\gamma_I) \frac{z_{IK}}{z_{IJ} z_{JK}}
\, c_1 + \ldots \right)^{(J)} + 
\langle\Sigma| \left( \ldots \right)^{(4)}
\nonumber \\
0 &=& \langle\Sigma| \left( c_{-2} + 
2 \rho_4^3 \left( 4\beta_4^3-6\beta_4\gamma_4+2\delta_4-
(1+\xi)(\gamma_4-\beta_4^2) \right) \, c_1 +
\ldots \right)^{(4)}
\nonumber \\
&+& \sum_{I=1}^3 \langle\Sigma| \left( 
2 (-1)^I \rho_4^2 \rho_I (1-\delta_{I1}) 
\frac{\beta_4^2-\gamma_4}{1-\xi} \, c_1 + 
\ldots \right)^{(I)} \,.
\label{CLc-2} \een

\subsection{An example}
We want here to give a simple but nontrivial example of a quartic 
correlator computation that uses some of the above conservation laws. Let us 
take one field of level four and one field of level six (see Section 
\ref{s-results} for the list of fields and their notation). We choose
\ben
|\Psi_4\rangle &=& c_{-1} \bar{c}_{-1} |0\rangle \nonumber \\
|\Psi_{12}\rangle &=& c_{-2} \bar{c}_{-2} |0\rangle \,.
\een
And we want to calculate the 
quartic amplitude of $\Psi_4$, $\Psi_{12}$, and two tachyons.
\be
\left\{T,\Psi_4,\Psi_{12},T\right\} = \frac{1}{\pi} \int_{{\cal V}_{0,4}} 
dx \wedge dy \langle \Sigma| {\cal B} {\cal B}^\star |T\rangle 
|\Psi_4\rangle |\Psi_{12}\rangle |T\rangle \,.
\label{TABT} \ee
where the antighost insertions are given by (\ref{BB}) and (\ref{Bim}). 
We find 
\ben
\langle \Sigma| {\cal B} {\cal B}^\star |T\rangle 
|\Psi_4\rangle |\Psi_{12}\rangle |T\rangle &=& -2 \left(B_1^2 \bar{B_1^2} - 
C_1^2 \bar{C_1^2} \right) \, \langle c_1, 1, c_{-2}, c_1 \rangle_o 
\langle \bar{c}_1, 1, \bar{c}_{-2}, \bar{c}_1 \rangle_o 
\nonumber \\
&& -2 \left(B_2^3 \bar{B_2^3} - 
C_2^3 \bar{C_2^3} \right) \, \langle c_1, c_{-1}, 1, c_1 \rangle_o 
\langle \bar{c}_1, \bar{c}_{-1}, 1, \bar{c}_1 \rangle_o 
\nonumber \\
&& +2 \left(B_1^2 \bar{B_2^3} - 
C_1^2 \bar{C_2^3} \right) \, \langle c_1, 1, c_{-2}, c_1 \rangle_o 
\langle \bar{c}_1, \bar{c}_{-1}, 1, \bar{c}_1 \rangle_o 
\nonumber \\
&& +2 \left(B_2^3 \bar{B_1^2} - 
C_2^3 \bar{C_1^2} \right) \, \langle c_1, c_{-1}, 1, c_1 \rangle_o 
\langle \bar{c}_1, 1, \bar{c}_{-2}, \bar{c}_1 \rangle_o \,.
\een
We therefore need to compute the two open correlators 
$\langle c_1, c_{-1}, 1, c_1 \rangle_o$ and 
$\langle c_1, 1, c_{-2}, c_1 \rangle_o$, on the 
four-punctured sphere $\Sigma$. To calculate the first one, 
we use the conservation laws (\ref{CLc-1}) to exchange the 
$c_{-1}$ on the second puncture for a $c_1$ on the second puncture 
and a $c_1$ on the third puncture. Namely
\ben
\langle c_1, c_{-1}, 1, c_1 \rangle_o &=& 
-\rho_2^2 \left( -4 \beta_2^2 + \beta_2 s_2 + 3 \gamma_2 - q_2 \right) 
\, \langle c_1, c_1, 1, c_1 \rangle_o \nonumber \\
&& + \rho_2 \rho_3 \frac{1}{\xi (1-\xi)} 
\left(\frac{1}{1-\xi} + \beta_2 - s_2 \right) \, 
\langle c_1, 1, c_1, c_1 \rangle_o \nonumber \\
&=& \frac{\rho_2}{\rho_1 \rho_4} \left( 4 \beta_2^2 
- \beta_2 - 3 \gamma_2 \right) \,.
\een
Similarly, we use the conservation laws for $c_{-2}$ (\ref{CLc-2}) 
to compute the second correlator by exchanging the $c_{-2}$ on 
the third puncture for a $c_1$ on the third puncture and a $c_1$ on 
the second puncture. We find
\ben
\langle c_1, 1, c_{-2}, c_1 \rangle_o &=& 
-2 \rho_3^3 \left( 4 \beta_3^3 - 6 \beta_3 \gamma_3 
+ 2 \delta_3 + (\gamma_3-\beta_3^2) s_3 \right) \, 
\langle c_1, 1, c_1, c_1 \rangle_o \nonumber \\
&& - 2 \rho_3^2 \rho_2 \left(\beta_3^2 - \gamma_3 \right) 
\frac{\xi}{\xi -1} \, \langle c_1, c_1, 1, c_1 \rangle_o
\nonumber \\
&=& -2 \frac{\rho_3^2}{\rho_1 \rho_4} \left( \xi \left( 
4 \beta_3^2 - 6 \beta_3 \gamma_3 + 2 \delta_3 \right) 
+ \gamma_3 - \beta_3^2 \right) \,.
\een
The integral in (\ref{TABT}) can then by expressed as an integral on
the region ${\cal A}$ (see (\ref{integration})) as explained in
\cite{dilaton}, and the numerical integration on ${\cal A}$ can be done by using the
fits given in \cite{quartic}.

\sectiono{The results}
\label{s-results}

\subsection*{The string field}

We start this section by writing the components of the string
field. We recall the string field up to level four, and compare our
notation with the one in \cite{vacuum}. Then we list all the fields at
level six. For level eight and ten, we describe a simple way to write
down all the closed fields from open fields of all ghost numbers.

\paragraph{}
We will write the string field in terms of components $\psi_i$ depending 
on one index.
\be
|\Psi\rangle = \sum_{i\geq1}\psi_i|\Psi_i\rangle \,.
\ee
The first field $|\Psi_1\rangle$ is the only field of level zero, namely the tachyon
\be
|\Psi_1\rangle = c_1 \bar{c}_1 \vac \,.
\ee
Then $|\Psi_2\rangle$ is the field of level two, the dilaton
\be
|\Psi_2\rangle = \left(c_1 c_{-1} - \bar{c}_1 \bar{c}_{-1}\right) \vac \,.
\ee
Before going further, it is good to introduce a way of listing the closed fields 
in a relatively simple manner. The elementary closed fields $|\Psi_k\rangle$ can 
be written 
\be
|\Psi_k\rangle = \left({\cal O}_{k_1} {\cal O}_{k_2}^\star - 
{\cal O}_{k_1}^\star {\cal O}_{k_2} \right) \vac \,,
\label{Psik} \ee
where ${\cal O}_{k_{1,2}}$ are products of left-moving oscillators. The $\star$ 
conjugation was defined in \cite{vacuum} on closed fields, here it simply 
changes all left-moving oscillators to right-moving oscillators without changing 
their order. Note that the expression (\ref{Psik}) is invariant under world-sheet parity 
${\cal P}$, whose action is
\be
{\cal P} \Psi = - \Psi^\star \,.
\ee
Indeed, it was shown in \cite{vacuum} that we may consistently
restrict the string field to have ${\cal P}$-eigenvalue one.  Let us
look at the open string states ${\cal O}_{k_1}\vac$ and ${\cal
O}_{k_2}\vac$. Because the closed string state must satisfy
$\left(L_0-\bar{L}_0\right) |\Psi_k\rangle = 0$, these two open string
states must have the same level $L$. Moreover, their ghost numbers
must add to two. If we write an open string state of level $L$ and
ghost number $G$ as $|L,G,i\rangle$, where $i$ is an index running
from one to the number $n_{L,G}$ of such open states, we can write
\be
|\Psi_k\rangle = |L_k, G_k, i_k\rangle \otimes |L_k, 2-G_k, j_k\rangle^\star - 
|L_k, G_k, i_k\rangle^\star \otimes |L_k, 2-G_k, j_k\rangle \,.
\label{open_to_closed} \ee
The definition of the $\star$-conjugation has been trivially extended
here, its action on a left-moving open string state is a right-moving
open-string state. We list the open string states $|L,G,i\rangle$ in
Table \ref{open1} for $L=0,1,2,3$ and in Table \ref{open2} for
$L=4,5$.
\begin{table}[!ht]
\begin{center}
\renewcommand\arraystretch{1.5}
\vskip 0.1in
\begin{tabular}{|c||c|l|c|}
  \hline
  $L$ & $G$ & open string states $|L,G,i\rangle$, $i=1,\ldots,n_{L,G}$ & $n_{L,G}$ \\
  \hline \hline
  $0$ & $1$ & $c_1 |0\rangle$ & $1$ \\
  \hline \hline
  $1$ & $0$ & $|0\rangle$ & $1$ \\
  \cline{2-4}
  & $2$ & $c_{-1} c_1|0\rangle$ & $1$ \\
  \hline \hline
  $2$ & $0$ & $b_{-2}c_1 |0\rangle$ & $1$ \\
  \cline{2-4}
  & $1$ & $c_{-1}|0\rangle$, \ $L_{-2} c_1 |0\rangle$ & $2$ \\
  \cline{2-4}
  & $2$ & $c_{-2} c_1 |0\rangle$ & $1$ \\
  \hline \hline
  $3$ & $-1$ & $b_{-2}\vac$ & $1$ \\
  \cline{2-4}
  & $0$ & $L_{-2}\vac$, \ $b_{-3}c_1\vac$ & $2$ \\
  \cline{2-4}
  & $1$ & $c_{-2}\vac$, \ $L_{-3}c_1\vac$, \ $b_{-2}c_{-1}c_1\vac$ & $3$ \\
  \cline{2-4}
  & $2$ & $c_{-3}c_1\vac$, \ $L_{-2}c_{-1}c_1\vac$ & $2$ \\
  \cline{2-4}
  & $3$ & $c_{-2}c_{-1}c_1\vac$ & $1$ \\
  \hline 
\end{tabular}
\caption{\footnotesize{The open string fields of level $L$ and ghost
number $G$} for levels 0 to 3.}
\label{open1}
\end{center}
\end{table}

\paragraph{}
Given these tables, it is now straightforward to write down all closed
fields at level $L$.  As a preliminary we see from the construction
(\ref{open_to_closed}) and from the fact that $n_{L,G}=n_{L,2-G}$,
that the number $N_{L}$ of closed string states at level $L$ is
\be
N_{L} = \sum_{G=2}^\infty n_{L/2,G}^2 + \frac{1}{2} n_{L/2,1} 
\left(n_{L/2,1}+1 \right) \,.
\label{NL} \ee
We list in Table \ref{NLTable}, the numbers $N_L$ for $L$ up to 24.
\begin{table}[!ht]
\begin{center}
\renewcommand\arraystretch{1.5}
\vskip 0.1in
\begin{tabular}{|c||c|c|c|c|c|c|c|c|c|c|c|c|c|}
\hline
$L$ & $0$ & $2$ & $4$ & $6$ & $8$ & $10$ & $12$ & $14$ & $16$ & $18$ & 
$20$ & $22$ & $24$ \\
\hline
$N_L$ & $1$ & $1$ & $4$ & $11$ & $38$ & $103$ & $314$ & $807$ & $2148$ & 
$5282$ & $12872$ & $29792$ & $68526$ \\
\hline
\end{tabular}
\caption{\footnotesize{The numbers of closed string states $N_L$ at level $L$.}}
\label{NLTable}
\end{center}
\end{table}
In this paper we shall limit ourselves to level 10, the computational
limit of our codes.

\paragraph{}
We can now continue to list the closed states. At level four, we read
from Table \ref{open1}
\ben
|\Psi_3\rangle &=& \left(b_{-2}c_1\bar{c}_{-2}\bar{c}_1 - 
\bar{b}_{-2}\bar{c}_1 c_{-2} c_1 \right) \vac \nonumber \\
|\Psi_4\rangle &=& c_{-1} \bar{c}_{-1} \vac \nonumber \\
|\Psi_5\rangle &=& L_{-2}c_1 \bar{L}_{-2} \bar{c}_1 \vac \nonumber \\
|\Psi_6\rangle &=& \left( c_{-1} \bar{L}_{-2} \bar{c}_1 - 
\bar{c}_{-1} L_{-2} c_1 \right) \vac \nonumber \,.
\een
So our fields $\psi_i$ up to level 4 are related to the fields of \cite{vacuum} by 
\be
\psi_1 = t \ , \quad \psi_2 = d \ , \quad \psi_3 = g_1 \ , \quad \psi_4 = f_1 
\ , \quad \psi_5 = f_2 \ , \quad \psi_6 = f_3 \,.
\ee
In order to facilitate comparisons, we will keep the names $t$, $d$, $g_1$, $f_1$, 
$f_2$ and $f_3$ for these fields. At level six, we have
\ben
|\Psi_7\rangle &=& \left(b_{-2} \bar{c}_{-2} \bar{c}_{-1} \bar{c}_{1} -
\bar{b}_{-2} c_{-2} c_{-1} c_{1} \right) \vac \nonumber \\
|\Psi_8\rangle &=& \left(L_{-2} \bar{c}_{-3} \bar{c}_1 - 
\bar{L}_{-2} c_{-3} c_{1} \right) \vac \nonumber \\
|\Psi_9\rangle &=& L_{-2} \bar{L}_{-2} \left(\bar{c}_{-1}\bar{c}_1  - 
c_{-1} c_1 \right) \vac \nonumber \\
|\Psi_{10}\rangle &=& \left(b_{-3} c_{1} \bar{c}_{-3} \bar{c}_{1} -
\bar{b}_{-3} \bar{c}_{1} c_{-3} c_{1} \right) \vac \nonumber \\
|\Psi_{11}\rangle &=& \left( b_{-3} c_1 \bar{L}_{-2} \bar{c}_{-1} \bar{c}_{1} -
\bar{b}_{-3} \bar{c}_1 L_{-2} c_{-1} c_{1} \right) \vac \nonumber \\
|\Psi_{12}\rangle &=& c_{-2} \bar{c}_{-2} \vac \nonumber \\
|\Psi_{13}\rangle &=& L_{-3} c_1 \bar{L}_{-3} \bar{c}_1 \vac \nonumber \\
|\Psi_{14}\rangle &=& b_{-2} c_{-1} c_{1} \bar{b}_{-2} \bar{c}_{-1} \bar{c}_{1}
\vac \nonumber \\
|\Psi_{15}\rangle &=& \left( c_{-2}\bar{L}_{-3}\bar{c}_{1} - 
\bar{c}_{-2}L_{-3}c_{1} \right) \vac \nonumber \\
|\Psi_{16}\rangle &=& \left(c_{-2}\bar{b}_{-2}\bar{c}_{-1}\bar{c}_{1} - 
\bar{c}_{-2}b_{-2}c_{-1}c_{1} \right) \vac \nonumber \\
|\Psi_{17}\rangle &=& \left(L_{-3}c_{1}\bar{b}_{-2}\bar{c}_{-1}\bar{c}_{1}  - 
\bar{L}_{-3}\bar{c}_{1}b_{-2}c_{-1}c_1 \right) \vac \nonumber \,.
\een

\begin{table}[!ht]
\begin{center}
\renewcommand\arraystretch{1.5}
\vskip 0.1in
\begin{tabular}{|c||c|l|c|}
  \hline
  $L$ & $G$ & open string states $|L,G,i\rangle$, $i=1,\ldots,n_{L,G}$ & $n_{L,G}$ \\
  \hline \hline
  $4$ & $-1$ & $b_{-3}\vac$ & $1$ \\
  \cline{2-4}
  & $0$ & $L_{-3}\vac$, \ $L_{-2}b_{-2}c_1\vac$, \ $b_{-4}c_1\vac$, \ 
  $b_{-2}c_{-1}\vac$ & $4$ \\
  \cline{2-4}
  & $1$ & $L_{-4}c_1\vac$, \ $L_{-2}c_{-1}\vac$, \ $L_{-2}L_{-2}c_1\vac$, \ 
  $c_{-3}\vac$, \ $b_{-3}c_{-1}c_1\vac$, \ $b_{-2}c_{-2}c_1\vac$  & $6$ \\
  \cline{2-4}
  & $2$ & $L_{-2}c_{-2}c_1\vac$, \ $L_{-3}c_{-1}c_1\vac$, \ 
  $c_{-4}c_1\vac$, \ $c_{-2}c_{-1}\vac$ & $4$ \\
  \cline{2-4}
  & $3$ & $c_{-3}c_{-1}c_1\vac$ & $1$ \\
  \hline \hline
  $5$ & $-1$ & $b_{-4}\vac$, \ $L_{-2}b_{-2}\vac$, \ $b_{-3}b_{-2}c_1\vac$ & $3$ \\
  \cline{2-4}
  & $0$ & $L_{-4}\vac$, \ $L_{-3}b_{-2}c_1\vac$, \ $L_{-2}L_{-2}\vac$, \ 
  $L_{-2}b_{-3}c_1\vac$, \ $b_{-5}c_1\vac$, \ $b_{-3}c_{-1}\vac$, \ 
  $b_{-2}c_{-2}\vac$ & $7$ \\
  \cline{2-4}
  & $1$ & $L_{-5}c_1\vac$, \ $L_{-3}c_{-1}\vac$, \ $L_{-3}L_{-2}c_1\vac$, \ 
  $L_{-2}c_{-2}\vac$, \ $L_{-2}b_{-2}c_{-1}c_1\vac$, \ $c_{-4}\vac$, & $9$ \\ 
  & & $b_{-4}c_{-1}c_1\vac$, \ $b_{-3}c_{-2}c_1\vac$, \ $b_{-2}c_{-3}c_1\vac$ & \\
  \cline{2-4}
  & $2$ & $L_{-4}c_{-1}c_1\vac$, \ $L_{-3}c_{-2}c_1\vac$, \ 
  $L_{-2}c_{-3}c_1\vac$, \ $L_{-2}L_{-2}c_{-1}c_1\vac$, \ $c_{-5}c_1\vac$, & $7$ \\ 
  & & $c_{-3}c_{-1}\vac$, \ $b_{-2}c_{-2}c_{-1}c_1\vac$ & \\
  \cline{2-4}
  & $3$ & $L_{-2}c_{-2}c_{-1}c_1\vac$, \ $c_{-4}c_{-1}c_1\vac$, \ $c_{-3}c_{-2}c_1\vac$ & $3$ \\
  \hline
\end{tabular}
\caption{\footnotesize{The open string fields of level $L$ and ghost number $G$} for levels 4 and 5.}
\label{open2}
\end{center}
\end{table}

\paragraph{} 
For levels 8 and 10, we don't explicitly write the $38+103$ fields, but we refer 
to Table \ref{open2}, and we specify in which order we do the constructions 
(\ref{open_to_closed}). First we do 
$G=-\infty,\ldots,0$, $i=1,\ldots,n_{L/2,G}$, $j=1,\ldots,n_{L/2,G}$. Then $G=1$, 
$i=j=1,\ldots,n_{L/2,G}$. And finally $G=1$, $i=1,\ldots,n_{L/2,G}$, 
$j=i+1,\ldots,n_{L/2,G}$.

\subsection*{The vacuum}

We will consider two different truncation schemes $A$ and $B$. In the
scheme $A$ (which was used in \cite{vacuum}), we keep all the
fields up to some fixed level $L$ (which in this paper will be
$L=10$), and we progressively increase the interaction level $M$ of
the quartic potential, $M=0,2,,\ldots,10$. In the scheme $B$ we
progressively increase the maximal fields level $L$ (here $L=2$ to
$L=10$ and we do not consider fields of level higher than $L$), and
for each $L$ we take the full quartic potential, i.e. the one with
interaction level $M = 4L$ (this is similar to what is usually done in
cubic string field theory).

\paragraph{}
We start by giving the relevant quartic potentials that we
computed. The quadratic and cubic potentials with fields up to level
six are written down in Appendix \ref{quad-cub}. For the truncation
scheme $B$, we need to extend the notations of \cite{vacuum}. We
define $V_{L,M}^{(4)}$ to be the quartic potential at level $M$ only,
with fields of level up to $L$. We note that if $L>M$, we have
$V_{L,M}^{(4)} = V_{M,M}^{(4)}$.  We then define the total potential
to level $M$ with fields to level $L$
\be
{\mathbb V}_{L,M}^{(4)} \equiv {\mathbb V}_{L,3L}^{(3)} + 
\sum_{i=0}^{M/2}V_{L,2i}^{(4)} \,,
\ee
where ${\mathbb V}_{L,3L}^{(3)}$ is the complete quadratic and cubic potential
with fields up to level $L$. We note that, at the highest level that we are considering, 
$L=10$, we take ${\mathbb V}_{10,24}^{(3)}$ instead of ${\mathbb V}_{10,30}^{(3)}$. 
Indeed this last potential is too big for our
symbolic calculator, but
we emphasize that the difference in the results is minute, as
can be verified by comparing the results using, for example, ${\mathbb
V}_{8,20}^{(3)}$ and ${\mathbb V}_{8,24}^{(3)}$. Scheme $B$ would
require that we compute all potentials up to $V_{10,40}^{(4)}$, but
this computation would be impossible within a reasonable time with our
codes on a desktop computer. We are able to compute $V_{0,0}^{(4)}$,
$V_{2,2}^{(4)}$, $V_{4,4}^{(4)}$, $V_{6,6}^{(4)}$, $V_{8,8}^{(4)}$,
$V_{10,10}^{(4)}$, $V_{10,12}^{(4)}$, $V_{6,14}^{(4)}$,
$V_{6,16}^{(4)}$. We will see below that these potentials are already
enough to give a good picture of scheme $B$. Of course if $L' < L$,
the potential $V_{L',M}^{(4)}$ can be obtained from $V_{L,M}^{(4)}$
simply by deleting the terms with fields of level greater than $L'$.
The quadratic and cubic potentials with fields up to level six are shown 
in Appendix \ref{quad-cub}. Here are some of the aforementioned quartic 
potentials
\ben
\kappa^2 V_{0,0}^{(4)} &=& -3.017 \, t^4 \nonumber \crbig
\kappa^2 V_{2,2}^{(4)} &=& 3.872 \, t^3 d \nonumber \crbig
\kappa^2 V_{4,4}^{(4)} &=& 1.368\,d^2 t^2 - 0.4377\,f_1 t^3 - 56.26\,f_2 t^3 + 13.02\,f_3 t^3 + 
0.2725\,g_1 t^3 \nonumber \crbig
\kappa^2 V_{6,6}^{(4)} &=& -0.9528 \, td^3 + t^2 d \left(
5.049 \, g_1 + 2.385 \, f_1 + 49.09 \, f_2 - 20.14 \, f_3 \right) \nonumber \\
&& + t^3 \left( 1.678 \, \psi_8 + 16.36 \, \psi_9 + 0.5357 \, \psi_{10} + 5.034 \, \psi_{11} - 
0.1790 \, \psi_{12} \right. \nonumber \\
&& \left. \hspace{24pt} - 91.70 \, \psi_{13} - 0.7159 \, \psi_{14} + 8.255 \, \psi_{15} + 
0.7159 \, \psi_{16} - 16.51 \, \psi_{17} \right) \nonumber \crbig
\kappa^2 V_{8,8}^{(4)} &=& -0.1056 \, d^4 + t d^2 \left( -3.226 \,
g_1 + 0.2779 \, f_1 + 19.31 \, f_2 - 5.047 \, f_3 \right) \nonumber \\
&& + t^2 d \left(1.043 \, \psi_7 - 2.393 \, \psi_8 + 19.31 \, \psi_9 + 1.325
\, \psi_{10} - 7.180 \, \psi_{11} + 0.3375 \, \psi_{12} \right. \nonumber \\
&& \left. \hspace{24pt} + 98.84 \, \psi_{13} + 1.350 \, \psi_{14} 
- 12.69 \, \psi_{15} - 2.393 \, \psi_{16} + 25.38 \, \psi_{17} \right) \nonumber \\ 
&& + t^2 \left(3.816 \, g_1^2 + 0.6519 \, g_1 f_1 
+ 3.429 \, g_1 f_2 + 1.025 \, g_1 f_3 + 0.2566 \, f_1^2 \right. \nonumber \\ 
&& \left. \hspace{24pt} - 10.98 \, f_1 f_2 - 1.906 \, f_1 f_3 - 979.3 \, f_2^2 
+ 314.4 \, f_2 f_3 - 10.59 \, f_3^2 \right) \nonumber \\
&& +t^3 \left( - 1.872 \, \psi_{19} + 32.94 \, \psi_{20} + 0.7143 \, \psi_{21} + 1.711 \, \psi_{22} 
+ 1.143 \, \psi_{23} - 3.750 \, \psi_{24} \right. \nonumber \\
&& \left. \hspace{24pt} + 0.09003 \, \psi_{25} - 0.3521 \, \psi_{26} + 
0.1803 \, \psi_{27} + 2.854 \, \psi_{28} + 0.1263 \, \psi_{29} + 0.09024 \, \psi_{30} 
\right. \nonumber \\
&& \left. \hspace{24pt}
- 0.3518 \, \psi_{31} + 422.0 \, \psi_{32} + 0.0452 \, \psi_{33} + 0.2043 \, \psi_{34} 
- 212.9 \, \psi_{35} - 3.660 \, \psi_{36} \right. \nonumber \\
&& \left. \hspace{24pt} - 831.3 \, \psi_{37} - 0.04596 \, \psi_{38} 
- 0.4136 \, \psi_{39} - 0.1758 \, \psi_{40} + 39.28 \, \psi_{41} - 658.1 \, \psi_{42} 
\right. \nonumber \\
&& \left. \hspace{24pt} + 7.068 \, \psi_{43} - 21.20 \, \psi_{44} - 9.795 \, \psi_{45} 
+ 123.6 \, \psi_{46} 
- 0.3480 \, \psi_{47} + 1.044 \, \psi_{48} \right. \nonumber \\ 
&& \left. \hspace{24pt} + 0.01764 \, \psi_{49} + 10.34 \, \psi_{50} 
- 31.01 \, \psi_{51} - 5.997 \, \psi_{52} + 0.2757 \, \psi_{53} + 0.1697 \, \psi_{54} 
\right. \nonumber \\
&& \left. \hspace{24pt} - 0.5091 \, \psi_{55} \right) \,.
\een
The numerical coefficients are rounded to four significant digits,
corresponding to the precision that the fit of the quartic geometry
\cite{quartic} allows to reach. 

\subsection*{Scheme $A$}

In Table \ref{Atable} we show our results for the nonperturbative
minimum of the potential in the truncation scheme $A$. We also give
the vacuum expectation values of the tachyon, dilaton and fields of
level four.
\begin{table}[!ht]
\begin{center}
\renewcommand\arraystretch{1.5}
\vskip 0.1in
\begin{tabular}{|c|c|c|c|c|c|c|c|}
\hline
\hbox{Potential} & $t$ & $d$ & $f_1$ & $f_2$ & $f_3$ & $g_1$ & 
\hbox{\small Value of the potential} \\
\hline
$\mathbb{V}^{(3)}_{10,24}$ & $0.4392$ & $0$ & $-0.06836$ & $-0.009648$ & $-0.02748$ & $0$ & $-0.06394$ \\
\hline \hline
$\mathbb{V}^{(4)}_{10,0}$ & $--$ & $--$ & $--$ & $--$ & $--$ & $--$ & $--$ \\
\hline
$\mathbb{V}^{(4)}_{10,2}$ & $0.3182$ & $0.4955$ & $-0.08272$ & $-0.006138$ & $-0.02679$ & $-0.1039$ & $-0.05429$ \\
\hline
$\mathbb{V}^{(4)}_{10,4}$ & $0.2311$ & $0.4638$ & $-0.04815$ & $-0.001680$ & $-0.01338$ & $-0.07412$ & $-0.03207$ \\
\hline
$\mathbb{V}^{(4)}_{10,6}$ & $0.4016$ & $0.4261$ & $-0.1457$ & $-0.008684$ & $-0.04016$ & $-0.03602$ & $-0.06860$ \\
\hline    
$\mathbb{V}^{(4)}_{10,8}$ & $0.3194$ & $0.4268$ & $-0.1322$ & $-0.01145$ & $-0.04284$ & $-0.1051$ & $-0.05368$ \\
\hline
$\mathbb{V}^{(4)}_{10,10}$ & $0.2901$ & $0.4587$ & $-0.1046$ & $-0.007365$ & $-0.03376$ & $-0.1095$ & $-0.04933$ \\
\hline
\end{tabular}
\caption{\footnotesize{The value of the potential and the expectation
values of the first few fields at the nonperturbative vacuum in the
truncation scheme $A$.}}
\label{Atable}
\end{center}
\end{table}
The lines up to interaction level four are very similar to the results
of \cite{vacuum}, the small differences coming from the quadratic and
cubic interactions with fields of level higher than four; these are
clearly unimportant contributions and the results agree
qualitatively. Looking at the value of the potential, we see that
although, up to level four, it seemed to approach monotonically zero,
it is actually oscillating around a value of about $-0.05$. This
oscillation, which is also visible on the expectation values of the
fields, is quite strong and makes it difficult to draw an accurate
conclusion from this data.

\subsection*{Scheme $B$}

Here we want to look at the minimum of ${\mathbb V}_{L,4L}^{(4)}$ for
$L=2, \ldots,10$. As we have already said, we can't fully compute
these potentials for $L>4$. To remedy this, we are going to look at
the values of the potential at the minimum of ${\mathbb V}_{L,M}^{(4)}$
for fixed $L$ and all $M$ starting at two and up as far as we
can. This data is shown in the columns of Table \ref{Btable}.
\begin{table}[!ht]
\begin{center}
\renewcommand\arraystretch{1.5}
\vskip 0.1in
\begin{tabular}{|c||c|c|c|c|c|}
\hline
$M$ & $L=2$ & $L=4$ & $L=6$ & $L=8$ & $L=10$ \\
\hline \hline
$2$ & $-0.1002$ & $-0.05806$ & $-0.05822$ & $-0.05422$ & $-0.05429$ \\
\hline
$4$ & $-0.05071$ & $-0.03383$ & $-0.03402$ & $-0.03199$ & $-0.03207$ \\
\hline 
$6$ & $-0.08141$ & $-0.07194$ & $-0.07204$ & $-0.06850$ & $-0.06860$ \\
\hline
$8$ & $-0.08534$ & $-0.05834$ & $-0.05674$ & $-0.05367$ & $-0.05368$ \\
\hline
$10$ & $--$ & $-0.05178$ & $-0.05181$ & $-0.04928$ & $-0.04933$ \\
\hline
$12$ & $--$ & $-0.05509$ & $-0.05516$ & $-0.05210$ & $-0.05193$ \\
\hline
$14$ & $--$ & $-0.05437$ & $-0.05427$ & $--$ & $--$ \\ 
\hline
$16$ & $--$ & $-0.05442$ & $-0.05438$ & $--$ & $--$ \\
\hline \hline
$4L$ & $-0.0853$ & $-0.0544$ & $-0.0544$ & $-0.0514$ & $-0.0513$ \\
\hline
\end{tabular}
\caption{\footnotesize{The values of the potentials $\kappa^2 V_{L,M}$
at the vacuum, and the extrapolation of the value of $\kappa^2
V_{L,4L}$.}}
\label{Btable}
\end{center}
\end{table}
Looking at the longest complete data that we have, namely $L=4$, we
see that the value of the potential at the vacuum oscillates (except
from $M=6$ to $M=10$ where it always increases when $L \geq 4$) and
converges relatively fast. We are thus making the assumption that the final 
result is always between the last two available values and closer to the last one, 
namely
\be
\kappa^2 {\mathbb V}_{L,4L}^{(4)} \approx \alpha \, \kappa^2 {\mathbb V}_{L,Q}^{(4)} + 
(1-\alpha) \, \kappa^2 {\mathbb V}_{L,Q+2}^{(4)} \,,
\ee
with $0<\alpha<0.5$. And we are making the further assumption that
$\alpha$ doesn't depend much on $Q$ and $L$. Once $\alpha$ is
estimated, we should use the larger $Q$ possible in order to have an
accurate extrapolation. The value of $\alpha$ that would give the
right answer for $L=4$ (with $Q=14$) is approximately
$\alpha=0.2$. But if we assume that $\kappa^2 {\mathbb
V}_{6,24}^{(4)}$ should be between $\kappa^2 {\mathbb V}_{6,14}^{(4)}$
and $\kappa^2 {\mathbb V}_{6,16}^{(4)}$, we should rather take $\alpha
\approx 0.25$. So we take $\alpha = 0.25$. The extrapolation for $L=6$
with $Q=14$ is then $\kappa^2 {\mathbb V}_{6,24}^{(4)} \approx
-0.0544$. For $L=8$ and $L=10$ we take $Q=10$ and we find $\kappa^2
{\mathbb V}_{8,32}^{(4)} \approx -0.0514$ and $\kappa^2 {\mathbb
V}_{10,40}^{(4)} \approx -0.0513$. As a check that $\alpha$ doesn't
depend much on $Q$, taking $Q=10$ for $L=4$ would give $\kappa^2
{\mathbb V}_{4,16}^{(4)} \approx -0.05426$, not terribly bad. We list
the values of $\kappa^2 {\mathbb V}_{L,4L}^{(4)}$ with three
significant digits, in the last line of Table \ref{Btable}.

\paragraph{}
Now we would like to make a final extrapolation to estimate 
$\kappa^2 {\mathbb V}_{L,4L}^{(4)}$ as $L \rightarrow \infty$. Fits of the form
\be
\kappa^2 {\mathbb V}_{L,4L}^{(4)} = f_0 + \frac{f_1}{L^\gamma}
\ee
are in general working quite well in open as well as closed string
field theory. The exponent $\gamma$, usually an integer or
half-integer, must be guessed in some way, more or less heuristically.
Since our values for $L=4,6$ and $L=8,10$ are very similar, we feed
the fit with only the values at $L=2,6,10$. Leaving $\gamma$ free, we
find that these three values are perfectly fitted with $\gamma = 1.76$,
and we take this as indication that we should take $\gamma = 2$. With
this last fit we find, with two significant digits
\be
\boxed{\lim_{L \rightarrow \infty} \kappa^2 {\mathbb V}_{L,4L}^{(4)} \approx -0.050} \,.
\label{limvacuum} \ee
Although it is harder to make an extrapolation from the data of the 
scheme $A$ (Table \ref{Atable}), the value (\ref{limvacuum}) fits 
well with it, in particular it is between the last two values.

\paragraph{}
We can do similar extrapolations of the vacuum expectation values of
the tachyon and dilaton. For the tachyon we obtain an oscillation
pattern very similar to the one of the potential value, and we find
\be
t \approx 0.29 \,.
\ee
The values for the dilaton, however, do not follow the same
oscillating pattern and we are not able to evaluate a reliable
extrapolation for $L>4$. At $L=2$ and $L=4$ we find $d = 0.439$ and
$d=0.435$ respectively. Our best estimation based on those two values
is thus
\be
d \approx 0.43 \,.
\ee
These values are again compatible with the data from scheme $A$.

\sectiono{Conclusions and prospects}
\label{s-conclusions}

In this paper we have considered nonpolynomial closed string field
theory truncated at polynomial order four. We have then truncated the
string field to level ten and have studied the nonperturbative minimum
of the potential. In \cite{vacuum}, an investigation of the
low-energy effective action of the tachyon, dilaton and graviton of
closed bosonic string theory led to the suggestion that if CSFT has a
nonperturbative minimum, its action density should vanish. The results
of the present paper do not support this supposition at quartic
order. Instead, we find that the quartic potential has a minimum with
height $-0.050$.

\paragraph{}
The question that we can ask now, is how the result (\ref{limvacuum})
changes as we include higher order terms in the action (i.e. quintic
term, sixtic term, etc...). In a separate paper \cite{quintic} one of
us has computed the five-tachyon contact term. Other quintic terms of 
higher level will follow \cite{Moe-prog}, but we want here to already see 
how the results change if we include the $t^5$ term in the potential. 
In our normalization we have \cite{quintic}
$$
\kappa^2 V_{0,0}^{(5)} = 9.924 \, t^5 \,.
$$ 
Since the tachyon expectation value is positive at the vacuum, we
expect this term to increase the value of the potential at the
minimum. We make the definition
$$
{\mathbb V}_{L,M}^{(4,t^5)} \equiv {\mathbb V}_{L,M}^{(4)} + V_{0,0}^{(5)} \,,
$$
and repeat our analysis in the truncation scheme $A$.
\begin{table}[!ht]
\begin{center}
\renewcommand\arraystretch{1.5}
\vskip 0.1in
\begin{tabular}{|c|c|c|c|c|c|c|c|}
\hline
\hbox{Potential} & $t$ & $d$ & $f_1$ & $f_2$ & $f_3$ & $g_1$ & 
\hbox{\small Value of the potential} \\
\hline
$\mathbb{V}^{(4,t^5)}_{10,0}$ & $0.3321$ & $0$ & $-0.03949$ & $-0.005976$ & $-0.01620$ & $0$ & $-0.05094$ \\
\hline
$\mathbb{V}^{(4,t^5)}_{10,2}$ & $0.2612$ & $0.2650$ & $-0.03506$ & $-0.003927$ & $-0.01285$ & $-0.04436$ & $-0.03380$ \\
\hline
$\mathbb{V}^{(4,t^5)}_{10,4}$ & $0.2187$ & $0.3460$ & $-0.03135$ & $-0.001509$ & $-0.009119$ & $-0.05061$ & $-0.02630$ \\
\hline
$\mathbb{V}^{(4,t^5)}_{10,6}$ & $0.2666$ & $0.2156$ & $-0.04480$ & $-0.003522$ & $-0.01353$ & $-0.01968$ & $-0.03370$ \\
\hline    
$\mathbb{V}^{(4,t^5)}_{10,8}$ & $0.2599$ & $0.2359$ & $-0.05041$ & $-0.004857$ & $-0.01657$ & $-0.03693$ & $-0.03276$ \\
\hline
$\mathbb{V}^{(4,t^5)}_{10,10}$ & $0.2570$ & $0.2479$ & $-0.04777$ & $-0.004227$ & $-0.01562$ & $-0.03966$ & $-0.03243$ \\
\hline
\end{tabular}
\caption{\footnotesize{The results of the truncation scheme $A$ with the term $t^5$ included.}}
\label{t5Atable}
\end{center}
\end{table}
We find, as expected, that all values of the potential are shallower. But we also note
that the oscillations are less strong than in Table \ref{Atable}; that
might be a sign that the results of level truncation will be improved
when we include the quintic term, and that this procedure of truncating the 
action order by order is convergent. We emphasize however that quintic terms 
of higher level are necessary to reach any conclusion.

\paragraph{}
The conclusion that we can make at this point, is that at quartic
order, the vacuum has a nonzero depth. It is possible that the higher
orders contributions are important enough to make this depth converge
to zero. It is also possible that the vacuum has a nonzero depth,
close to what we find at quartic order. In this last case, it will be very
interesting to try to understand what is this vacuum. Hopefully, the
upcoming calculation at quintic order will make it possible to decide
which one of the two alternatives is the right one.

\section*{Acknowledgments}

We would like to thank Barton Zwiebach for helpful discussions and
suggestions on the manuscript. N.~M. is supported by an "EC"
fellowship within the framework of the "Marie Curie Research Training
Network" Programme, Contract no. MRTN-CT-2004-503369.

\appendix

\sectiono{The quadratic and cubic potentials with fields of level up to six}
\label{quad-cub}
\small

In this appendix we want to write the potential ${\mathbb V}_{L,3L}^{(3)}$ with 
the fields level $L=6$. It is decomposed in terms of quadratic potentials
$V_M^{(2)}$ and cubic potentials $V_M^{(3)}$ at level $M$.
\be
{\mathbb V}_{6,18}^{(3)} = 
V_0^{(2)} + V_8^{(2)} + V_{12}^{(2)} + V_0^{(3)} + 
V_4^{(3)} + V_6^{(3)} + V_8^{(3)} + V_{10}^{(3)} + V_{12}^{(3)}
 + V_{14}^{(3)} + V_{16}^{(3)} + V_{18}^{(3)} \,.
\ee
For the quadratic potentials we have
\ben
\kappa^2 V_0^{(2)} &=& -t^2 \\
\kappa^2 V_8^{(2)} &=& f_1^2+169 f_2^2-26 f_3^2-2 g_1^2 \\
\kappa^2 V_{12}^{(2)} &=& 4 \psi_7^2-676 \psi_9^2-4
\psi_{10}^2+5408 \psi_{13}^2+4 \psi_{16}^2-104 \psi_8 \psi_{11} +4
\psi_{12} \psi_{14} +416 \psi_{15} \psi_{17} \,.
\een
And the cubic potentials are
\ben \kappa^2 V_0^{(3)} \hspace{-6pt} &=& \hspace{-6pt} \frac{6561
t^3}{4096} \\ \kappa^2 V_4^{(3)} \hspace{-6pt} &=& \hspace{-6pt} -\frac{27 t
d^2}{32}+\frac{3267 f_1 t^2}{4096}+\frac{114075 f_2
t^2}{4096}-\frac{19305 f_3 t^2}{2048} \\ \kappa^2 V_6^{(3)}
\hspace{-6pt} &=& \hspace{-6pt} -\frac{25}{8} d g_1 t \\ \kappa^2
V_8^{(3)} \hspace{-6pt} &=& \hspace{-6pt} -\frac{f_1
d^2}{96}-\frac{4225 f_2 d^2}{864}+\frac{65 f_3
d^2}{144}+\frac{325}{432} t \psi_8 d-\frac{4225}{432} t \psi_9
d-\frac{25}{144} t \psi_{10} d+\frac{325}{144} t \psi_{11} d+\frac{361
f_1^2 t}{12288}\nonumber \\ 
&& \hspace{-6pt} +\frac{57047809 f_2^2 t}{110592}+\frac{470873 f_3^2
t}{27648}-\frac{49 g_1^2 t}{24}+\frac{511225 f_1 f_2
t}{55296}-\frac{13585 f_1 f_3 t}{9216}-\frac{5400395 f_2 f_3 t}{27648}
\\ \kappa^2 V_{10}^{(3)} \hspace{-6pt} &=& \hspace{-6pt}
-\frac{400}{729} \psi_7 d^2+\frac{50}{729} \psi_{12}
d^2+\frac{200}{729} \psi_{14} d^2+\frac{200}{729} \psi_{16}
d^2-\frac{9025 f_1 g_1 d}{5832}-\frac{105625 f_2 g_1
d}{5832}+\frac{30875 f_3 g_1 d}{2916} \nonumber \\ 
&& \hspace{-6pt} +\frac{6175 g_1 t
\psi_8}{1944}-\frac{105625 g_1 t \psi_9}{5832}-\frac{361}{216} g_1 t
\psi_{10} +\frac{6175}{648} g_1 t \psi_{11} +\frac{50}{729} f_1 t
\psi_{12} +\frac{346112}{729} f_2 t \psi_{13} +\frac{200}{729} f_1 t
\psi_{14} \nonumber \\
&& \hspace{-6pt} -\frac{8320}{729} f_3 t \psi_{15}-\frac{200}{729} f_1 t
\psi_{16} +\frac{16640}{729} f_3 t \psi_{17} \\ \kappa^2 V_{12}^{(3)}
\hspace{-6pt} &=& \hspace{-6pt} \frac{f_1^3}{4096}+\frac{1525225 f_2
f_1^2}{8957952}-\frac{1235 f_3 f_1^2}{55296}+\frac{6902784889 f_2^2
f_1}{80621568}+\frac{1884233 f_3^2 f_1}{2239488}-\frac{961 g_1^2
f_1}{157464}-\frac{102607505 f_2 f_3 f_1}{6718464} \nonumber \\ 
&& \hspace{-6pt} +\frac{325 d \psi_8
f_1}{34992}-\frac{4225 d \psi_9 f_1}{34992}-\frac{25 d \psi_{10}
f_1}{11664}+\frac{325 d \psi_{11} f_1}{11664}+\frac{74181603769
f_2^3}{26873856}-\frac{31167227 f_3^3}{3359232}\nonumber \\ 
&& \hspace{-6pt} +\frac{4965049817 f_2
f_3^2}{20155392} -\frac{207025 f_2 g_1^2}{17496}+\frac{14105 f_3
g_1^2}{26244}+\frac{128 t \psi_7^2}{19683}-\frac{105625 t
\psi_8^2}{629856}-\frac{57047809 t \psi_9^2}{629856}\nonumber \\ 
&& \hspace{-6pt} -\frac{1207801 t
\psi_{10}^2}{629856}-\frac{105625 t \psi_{11}^2}{69984} +\frac{625 t
\psi_{12}^2}{19683}+\frac{44302336 t \psi_{13}^2}{19683}+\frac{10000 t
\psi_{14}^2}{19683}-\frac{332800 t \psi_{15}^2}{19683}+\frac{8528 t
\psi_{16}^2}{19683}\nonumber \\ 
&& \hspace{-6pt} -\frac{1331200 t
\psi_{17}^2}{19683}-\frac{22628735129 f_2^2 f_3}{13436928} -\frac{33856
d g_1 \psi_7}{19683}+\frac{2454725 d f_2 \psi_8}{314928}-\frac{9815 d
f_3 \psi_8}{17496}-\frac{57047809 d f_2 \psi_9}{314928}\nonumber \\ 
&& \hspace{-6pt} +\frac{490945 d
f_3 \psi_9}{52488}+\frac{2454725 t \psi_8 \psi_9}{314928} -\frac{105625
d f_2 \psi_{10}}{104976}+\frac{1625 d f_3
\psi_{10}}{17496}+\frac{357175 t \psi_8
\psi_{10}}{314928}-\frac{105625 t \psi_9
\psi_{10}}{104976}\nonumber \\
&& \hspace{-6pt} +\frac{2454725 d f_2 \psi_{11}}{104976}-\frac{9815 d
f_3 \psi_{11}}{5832} -\frac{8300747 t \psi_8
\psi_{11}}{314928}+\frac{2454725 t \psi_9
\psi_{11}}{104976}+\frac{357175 t \psi_{10}
\psi_{11}}{104976}+\frac{1400 d g_1 \psi_{12}}{6561}\nonumber \\ 
&& \hspace{-6pt} +\frac{5600 d g_1
\psi_{14}}{6561}+\frac{392 t \psi_{12} \psi_{14}}{2187} +\frac{17056 d
g_1 \psi_{16}}{19683}-\frac{1400 t \psi_{12}
\psi_{16}}{6561}-\frac{5600 t \psi_{14} \psi_{16}}{6561}+\frac{372736
t \psi_{15} \psi_{17}}{6561} \\ \kappa^2 V_{14}^{(3)} \hspace{-6pt}
&=& \hspace{-6pt} -\frac{211250 \psi_{12}
f_3^2}{531441}-\frac{41879552 \psi_{13} f_3^2}{531441}-\frac{845000
\psi_{14} f_3^2}{531441}+\frac{5948800 \psi_{15}
f_3^2}{531441}+\frac{845000 \psi_{16} f_3^2}{531441}\nonumber \\
&& \hspace{-6pt} -\frac{11897600
\psi_{17} f_3^2}{531441}-\frac{27055015 g_1 \psi_8
f_3}{2125764}+\frac{233198875 g_1 \psi_9 f_3}{2125764}+\frac{1021345
g_1 \psi_{10} f_3}{236196}-\frac{27055015 g_1 \psi_{11}
f_3}{708588} \nonumber \\ 
&& \hspace{-6pt} -\frac{78400 g_1^2 \psi_7}{59049}+\frac{5106725 f_1 g_1
\psi_8}{4251528}+\frac{46639775 f_2 g_1 \psi_8}{1417176}+\frac{426400
d \psi_7 \psi_8}{531441}-\frac{38130625 f_1 g_1
\psi_9}{4251528} \nonumber \\
&& \hspace{-6pt} -\frac{1426195225 f_2 g_1
\psi_9}{4251528}-\frac{3380000 d \psi_7 \psi_9}{531441}-\frac{683929
f_1 g_1 \psi_{10}}{1417176}-\frac{1525225 f_2 g_1
\psi_{10}}{157464}-\frac{53792 d \psi_7
\psi_{10}}{177147} \nonumber \\ 
&& \hspace{-6pt} +\frac{5106725 f_1 g_1
\psi_{11}}{1417176}+\frac{46639775 f_2 g_1
\psi_{11}}{472392}+\frac{426400 d \psi_7 \psi_{11}}{177147}+\frac{9800
g_1^2 \psi_{12}}{59049}+\frac{211250 f_1 f_2
\psi_{12}}{531441} \nonumber \\ 
&& \hspace{-6pt} -\frac{32500 d \psi_8
\psi_{12}}{531441}+\frac{422500 d \psi_9 \psi_{12}}{531441}+\frac{1900
d \psi_{10} \psi_{12}}{59049}-\frac{24700 d \psi_{11}
\psi_{12}}{59049}+\frac{41879552 f_1 f_2
\psi_{13}}{531441} \nonumber \\ 
&& \hspace{-6pt} -\frac{44302336 d \psi_9
\psi_{13}}{531441}+\frac{39200 g_1^2 \psi_{14}}{59049}+\frac{845000
f_1 f_2 \psi_{14}}{531441}-\frac{98800 d \psi_8
\psi_{14}}{177147}+\frac{1690000 d \psi_9
\psi_{14}}{531441} \nonumber \\ 
&& \hspace{-6pt} +\frac{7600 d \psi_{10}
\psi_{14}}{59049}-\frac{130000 d \psi_{11}
\psi_{14}}{177147}-\frac{5948800 f_1 f_2
\psi_{15}}{531441}-\frac{332800 d \psi_8
\psi_{15}}{531441}+\frac{252928 d \psi_{11}
\psi_{15}}{59049} \nonumber \\ 
&& \hspace{-6pt} +\frac{39200 g_1^2 \psi_{16}}{59049}-\frac{845000 f_1
f_2 \psi_{16}}{531441}-\frac{213200 d \psi_8
\psi_{16}}{531441}+\frac{1690000 d \psi_9
\psi_{16}}{531441}+\frac{30992 d \psi_{10}
\psi_{16}}{177147} \nonumber \\
&& \hspace{-6pt} -\frac{213200 d \psi_{11}
\psi_{16}}{177147}+\frac{11897600 f_1 f_2
\psi_{17}}{531441}-\frac{505856 d \psi_8
\psi_{17}}{177147}+\frac{665600 d \psi_{11} \psi_{17}}{177147} \\
\kappa^2 V_{16}^{(3)} \hspace{-6pt} &=& \hspace{-6pt} \frac{5274752
f_1 \psi_7^2}{14348907}+\frac{540800 f_2
\psi_7^2}{14348907}-\frac{3377920 f_3
\psi_7^2}{14348907}+\frac{6219200 g_1 \psi_8
\psi_7}{4782969}-\frac{143041600 g_1 \psi_9
\psi_7}{14348907} \nonumber \\ 
&& \hspace{-6pt} -\frac{270400 g_1 \psi_{10}
\psi_7}{531441}+\frac{6219200 g_1 \psi_{11}
\psi_7}{1594323}-\frac{105625 f_1 \psi_8^2}{51018336}-\frac{1426195225
f_2 \psi_8^2}{459165024}+\frac{12273625 f_3
\psi_8^2}{76527504} \nonumber \\
&& \hspace{-6pt} -\frac{57047809 f_1
\psi_9^2}{51018336}-\frac{74181603769 f_2
\psi_9^2}{51018336}+\frac{2057157739 f_3
\psi_9^2}{25509168}-\frac{131997121 f_1
\psi_{10}^2}{459165024}-\frac{5102959225 f_2
\psi_{10}^2}{459165024} \nonumber \\ 
&& \hspace{-6pt} +\frac{820716715 f_3
\psi_{10}^2}{229582512}-\frac{105625 f_1
\psi_{11}^2}{5668704}-\frac{1426195225 f_2
\psi_{11}^2}{51018336}+\frac{12273625 f_3
\psi_{11}^2}{8503056}+\frac{625 f_1 \psi_{12}^2}{19683} \nonumber \\
&& \hspace{-6pt} +\frac{2640625
f_2 \psi_{12}^2}{14348907}-\frac{81250 f_3
\psi_{12}^2}{531441}+\frac{5360582656 f_1
\psi_{13}^2}{14348907}+\frac{15993143296 f_2
\psi_{13}^2}{1594323}-\frac{18518376448 f_3
\psi_{13}^2}{4782969} \nonumber \\
&& \hspace{-6pt} +\frac{10000 f_1
\psi_{14}^2}{19683}+\frac{42250000 f_2
\psi_{14}^2}{14348907}-\frac{1300000 f_3
\psi_{14}^2}{531441}-\frac{3660800 f_1
\psi_{15}^2}{531441}-\frac{411008000 f_2
\psi_{15}^2}{4782969} \nonumber \\
&& \hspace{-6pt} +\frac{750131200 f_3
\psi_{15}^2}{14348907}+\frac{3795152 f_1
\psi_{16}^2}{14348907}+\frac{36030800 f_2
\psi_{16}^2}{14348907}-\frac{4852640 f_3
\psi_{16}^2}{4782969}-\frac{14643200 f_1
\psi_{17}^2}{531441} \nonumber \\ 
&& \hspace{-6pt} -\frac{1644032000 f_2
\psi_{17}^2}{4782969}+\frac{3000524800 f_3
\psi_{17}^2}{14348907}+\frac{2454725 f_1 \psi_8
\psi_9}{25509168}+\frac{10285788695 f_2 \psi_8
\psi_9}{76527504} \nonumber \\
&& \hspace{-6pt} -\frac{275396485 f_3 \psi_8
\psi_9}{38263752}+\frac{3733925 f_1 \psi_8
\psi_{10}}{76527504}+\frac{2697742775 f_2 \psi_8
\psi_{10}}{229582512}-\frac{251765605 f_3 \psi_8
\psi_{10}}{114791256} \nonumber \\ 
&& \hspace{-6pt} -\frac{105625 f_1 \psi_9
\psi_{10}}{8503056}-\frac{1426195225 f_2 \psi_9
\psi_{10}}{76527504}+\frac{12273625 f_3 \psi_9
\psi_{10}}{12754584}-\frac{86776417 f_1 \psi_8
\psi_{11}}{76527504} \nonumber \\
&& \hspace{-6pt} -\frac{19456250905 f_2 \psi_8
\psi_{11}}{76527504}+\frac{1834363531 f_3 \psi_8
\psi_{11}}{38263752}+\frac{2454725 f_1 \psi_9
\psi_{11}}{8503056}+\frac{10285788695 f_2 \psi_9
\psi_{11}}{25509168} \nonumber \\ 
&& \hspace{-6pt} -\frac{275396485 f_3 \psi_9
\psi_{11}}{12754584}+\frac{3733925 f_1 \psi_{10}
\psi_{11}}{25509168}+\frac{2697742775 f_2 \psi_{10}
\psi_{11}}{76527504}-\frac{251765605 f_3 \psi_{10}
\psi_{11}}{38263752} \nonumber \\
&& \hspace{-6pt} -\frac{455000 g_1 \psi_8
\psi_{12}}{4782969}+\frac{5915000 g_1 \psi_9
\psi_{12}}{4782969}+\frac{26600 g_1 \psi_{10}
\psi_{12}}{531441}-\frac{345800 g_1 \psi_{11}
\psi_{12}}{531441}-\frac{2215116800 g_1 \psi_9
\psi_{13}}{14348907} \nonumber \\
&& \hspace{-6pt} -\frac{1383200 g_1 \psi_8
\psi_{14}}{1594323}+\frac{23660000 g_1 \psi_9
\psi_{14}}{4782969}+\frac{106400 g_1 \psi_{10}
\psi_{14}}{531441}-\frac{1820000 g_1 \psi_{11}
\psi_{14}}{1594323}+\frac{150152 f_1 \psi_{12}
\psi_{14}}{14348907} \nonumber \\
&& \hspace{-6pt} +\frac{1656200 f_2 \psi_{12}
\psi_{14}}{1594323}+\frac{997360 f_3 \psi_{12}
\psi_{14}}{4782969}+\frac{4659200 g_1 \psi_8
\psi_{15}}{4782969}+\frac{34611200 g_1 \psi_9
\psi_{15}}{14348907}-\frac{12646400 g_1 \psi_{11}
\psi_{15}}{1594323} \nonumber \\
&& \hspace{-6pt} -\frac{3234400 g_1 \psi_8
\psi_{16}}{4782969}+\frac{72061600 g_1 \psi_9
\psi_{16}}{14348907}+\frac{164000 g_1 \psi_{10}
\psi_{16}}{531441}-\frac{3234400 g_1 \psi_{11}
\psi_{16}}{1594323}+\frac{27400 f_1 \psi_{12}
\psi_{16}}{531441} \nonumber \\
&& \hspace{-6pt} -\frac{5915000 f_2 \psi_{12}
\psi_{16}}{4782969}+\frac{5590000 f_3 \psi_{12}
\psi_{16}}{14348907}+\frac{109600 f_1 \psi_{14}
\psi_{16}}{531441}-\frac{23660000 f_2 \psi_{14}
\psi_{16}}{4782969} \nonumber \\
&& \hspace{-6pt} +\frac{22360000 f_3 \psi_{14}
\psi_{16}}{14348907}+\frac{25292800 g_1 \psi_8
\psi_{17}}{4782969}-\frac{69222400 g_1 \psi_9
\psi_{17}}{14348907}-\frac{9318400 g_1 \psi_{11}
\psi_{17}}{1594323} \nonumber \\
&& \hspace{-6pt} -\frac{80244736 f_1 \psi_{15}
\psi_{17}}{14348907}+\frac{460328960 f_2 \psi_{15}
\psi_{17}}{1594323}-\frac{127901696 f_3 \psi_{15} \psi_{17}}{4782969}
\\ \kappa^2 V_{18}^{(3)} \hspace{-6pt} &=& \hspace{-6pt}
-\frac{99123200 \psi_{12} \psi_7^2}{387420489}-\frac{396492800
\psi_{14} \psi_7^2}{387420489}-\frac{396492800 \psi_{16}
\psi_7^2}{387420489}-\frac{57402800 \psi_8^2
\psi_7}{129140163}-\frac{22819123600 \psi_9^2
\psi_7}{387420489} \nonumber \\
&& \hspace{-6pt} -\frac{144400 \psi_{10}^2
\psi_7}{4782969}-\frac{57402800 \psi_{11}^2 \psi
_7}{14348907}+\frac{3220599200 \psi_8 \psi_9
\psi_7}{387420489}+\frac{8101600 \psi_8 \psi_{10}
\psi_7}{43046721}-\frac{227271200 \psi_9 \psi_{10}
\psi_7}{129140163} \nonumber \\
&& \hspace{-6pt} -\frac{227271200 \psi_8 \psi_{11}
\psi_7}{129140163}+\frac{3220599200 \psi_9 \psi_{11}
\psi_7}{129140163}+\frac{8101600 \psi_{10} \psi _{11}
\psi_7}{14348907}+\frac{5281250 \psi_8^2
\psi_{12}}{387420489} \nonumber \\
&& \hspace{-6pt} +\frac{2852390450 \psi_9^2
\psi_{12}}{387420489}+\frac{18050 \psi_{10}^2 \psi
_{12}}{4782969}+\frac{3050450 \psi_{11}^2
\psi_{12}}{4782969}-\frac{245472500 \psi_8 \psi_9
\psi_{12}}{387420489}-\frac{617500 \psi_8 \psi_{10} \psi
_{12}}{43046721} \nonumber \\
&& \hspace{-6pt} +\frac{8027500 \psi_9 \psi_{10}
\psi_{12}}{43046721}+\frac{14350700 \psi_8 \psi_{11}
\psi_{12}}{43046721}-\frac{186559100 \psi_9 \psi_{11}
\psi_{12}}{43046721}-\frac{469300 \psi_{10} \psi_{11}
\psi_{12}}{4782969} \nonumber \\
&& \hspace{-6pt} +\frac{553779200 \psi_8^2
\psi_{13}}{387420489}+\frac{553779200 \psi_{11}^2 \psi
_{13}}{43046721}-\frac{1107558400 \psi_9 \psi_{10}
\psi_{13}}{129140163}+\frac{12201800 \psi_8^2
\psi_{14}}{43046721} \nonumber \\
&& \hspace{-6pt} +\frac{11409561800 \psi_9^2 \psi
_{14}}{387420489}+\frac{72200 \psi_{10}^2
\psi_{14}}{4782969}+\frac{21125000 \psi_{11}^2
\psi_{14}}{43046721}-\frac{746236400 \psi_8 \psi_9 \psi
_{14}}{129140163}-\frac{1877200 \psi_8 \psi_{10}
\psi_{14}}{14348907} \nonumber \\
&& \hspace{-6pt} +\frac{32110000 \psi_9 \psi_{10}
\psi_{14}}{43046721}+\frac{57402800 \psi_8 \psi_{11}
\psi_{14}}{43046721}-\frac{981890000 \psi_9 \psi_{11}
\psi_{14}}{129140163}-\frac{2470000 \psi_{10} \psi_{11}
\psi_{14}}{14348907} \nonumber \\
&& \hspace{-6pt} +\frac{108160000 \psi_8^2
\psi_{15}}{387420489}-\frac{82201600 \psi_{11}^2
\psi_{15}}{14348907}-\frac{2513638400 \psi_8 \psi_9
\psi_{15}}{387420489}-\frac{6323200 \psi_8 \psi_{10}
\psi_{15}}{43046721} \nonumber \\
&& \hspace{-6pt} +\frac{138444800 \psi_9 \psi_{10}
\psi_{15}}{129140163}+\frac{2513638400 \psi_9 \psi_{11}
\psi_{15}}{129140163}+\frac{6323200 \psi_{10} \psi_{11}
\psi_{15}}{14348907}+\frac{16055000 \psi_8^2
\psi_{16}}{129140163} \nonumber \\
&& \hspace{-6pt} +\frac{11409561800 \psi_9^2
\psi_{16}}{387420489}+\frac{72200 \psi_{10}^2 \psi
_{16}}{4782969}+\frac{16055000 \psi_{11}^2
\psi_{16}}{14348907}-\frac{1610299600 \psi_8 \psi_9
\psi_{16}}{387420489}-\frac{4050800 \psi_8 \psi_{10} \psi
_{16}}{43046721} \nonumber \\
&& \hspace{-6pt} +\frac{130941200 \psi_9 \psi_{10}
\psi_{16}}{129140163}+\frac{57402800 \psi_8 \psi_{11}
\psi_{16}}{43046721}-\frac{1610299600 \psi_9 \psi _{11}
\psi_{16}}{129140163}-\frac{4050800 \psi_{10} \psi_{11}
\psi_{16}}{14348907} \nonumber \\
&& \hspace{-6pt} +\frac{164403200 \psi_8^2
\psi_{17}}{129140163}-\frac{216320000 \psi_{11}^2
\psi_{17}}{43046721}-\frac{5027276800 \psi_8 \psi_9
\psi_{17}}{387420489}-\frac{12646400 \psi_8 \psi_{10}
\psi_{17}}{43046721} \nonumber \\
&& \hspace{-6pt} -\frac{276889600 \psi_9 \psi _{10}
\psi_{17}}{129140163}+\frac{5027276800 \psi_9 \psi_{11}
\psi_{17}}{129140163}+\frac{12646400 \psi_{10} \psi_{11}
\psi_{17}}{14348907} \,. \een


\end{document}